\documentclass[epsfig]{elsart}
\usepackage{epsfig}

\begin{document}

\begin{frontmatter}

\title{Non-Gaussian Resistance Noise near Electrical Breakdown 
in Granular Materials}

\author{C. Pennetta\thanksref{a}\corauthref{*}}
\ead{cecilia.pennetta@unile.it}
\author{, E. Alfinito\thanksref{a}} 
\author{, L. Reggiani\thanksref{a},}
\author{S. Ruffo\thanksref{2}}

\corauth[*]{Corresponding author. Dipartimento di Ingegneria dell'Innovazione, 
Universit\`a di Lecce, Via Arnesano, 73100, Italy.}

\address[a]{INFM, National Nanotechnology Lab., Via Arnesano, 73100, Italy 
and Dip. di Ingegneria dell'Innovazione, 
Universit\`a di Lecce, Via Arnesano, 73100, Italy.}
\address[2]{CSDC and Dipartimento di Energetica ``Sergio Stecco'', 
\\Universit\`a di Firenze, Via S. Marta, 3, INFN and INFM, Firenze, 
50139, Italy.}

\thanks[1]{Partial support from the cofin-03 project ``Modelli e misure di
rumore in nanostrutture'' financed by Italian MIUR and from SPOT-NOSED  
project IST-2001-38899 of EC is gratefully acknowledged.}

\begin{abstract}
The distribution of resistance fluctuations of conducting thin films 
with granular structure near electrical breakdown is studied by 
numerical simulations. The film is modeled as a resistor network in a steady 
state determined by the competition between two biased processes, breaking and
recovery. Systems of different sizes and with different levels of internal 
disorder are considered. Sharp deviations from a Gaussian distribution are 
found near breakdown and the effect increases with the degree of internal 
disorder. However, we show that in general this non-Gaussianity is related to 
the finite size of the system and vanishes in the large size limit. 
Nevertheless, near the critical point of the conductor-insulator transition, 
deviations from Gaussianity persist when the size is increased and the 
distribution of resistance fluctuations is well fitted by the universal 
Bramwell-Holdsworth-Pinton distribution. 

\end{abstract}

\end{frontmatter}

\vspace{-1. cm}
\section{Introduction and Model}
\label{3}
\vspace{-0.8 cm}
Non-Gaussian distributions of several quantities characterizing the behavior 
of complex systems in non-equilibrium steady-states, have been evidenced 
since long time in many experiments \cite{weissman}. The general 
properties of these anomalous distributions and their link with other 
properties of non-equilibrium complex systems is far from being fully 
understood. Rather, the study of these distributions is nowadays attracting 
an increasing interest in the literature 
\cite{chakrabarti,bramwell_nat,bramwell_prl,bramwell_pre,racz,ausloos}. 
Here, we study the distribution of resistance fluctuations of a conducting
thin film with granular structure near electrical breakdown. Electrical 
breakdown phenomena typically occur in conductors stressed by high current 
densities and they consist in an irreversible and dramatic increase of  
resistivity of the material and thus they are associated with a
conductor-insulator transition 
\cite{hansen,sornette,stan_zap,bardhan,prl_fail}. In our study we make use of 
the Stationary Network Under Biased Percolation (SNUBP) model \cite{SNUBP}. 
This model provides a good description of many features associated with the 
electrical instability of composites materials \cite{bardhan} and with the
electromigration damage of metal lines \cite{pen_em}, two important classes of 
breakdown phenomena. The film is modeled as a resistor network which reaches
a steady state determined by the competition between two biased stochastic 
processes, breaking and recovery. Systems of different sizes and with 
different levels of internal disorder are considered, where the disorder is 
mainly related to the fraction of broken resistors within the network. The 
resistance and its fluctuations are calculated by Monte Carlo simulations 
which are performed under different stress conditions. 
Resistance fluctuations are found to deviate from Gaussianity near electrical 
breakdown. We analyze and discuss this non-Gaussianity in the spirit of the 
Bramwell-Holdsworth-Pinton (BHP) distribution, recently introduced in the
context of the study of highly correlated systems near criticality 
\cite{bramwell_nat,bramwell_prl}. In particular, we show that the deviations
from Gaussianity observed near the breakdown are related to finite size 
effects, and consequently vanish in the thermodynamic limit. Nevertheless, 
near the percolation critical point the non-Gaussianity persists in the large 
size limit and it is described by the universal BHP distribution. 

\vspace{-0.5 cm}
According to the SNUBP model\cite{SNUBP}, a conducting film with 
granular structure is described as a two-dimensional resistor network. 
Precisely, we consider a square-lattice of $N \times N$ resistors. This
lattice lies on an insulating substrate at a given temperature $T_0$, which
acts as a thermal bath. Each resistor can be in two different states: 
(i) regular, corresponding to a resistance $r_n =r_0[1+\alpha(T_n - T_0)]$ 
and (ii) broken, corresponding to the effectively ``infinite'' resistance, 
$r_{OP} = 10^9 r_n$ (in the following we call this state defect). Here, 
$\alpha$ is the temperature coefficient of the resistance and $T_n$ the local 
temperature. This latter is determined by taking into account Joule heating 
effects and thermal exchanges between neighbor resistors \cite{prl_fail}:
$T_{n}=T_{0} + A [ r_{n} i_{n}^{2} + (3/4N_{neig})
\sum  ( r_{l} i_{l}^2   - r_n i_n^2)]$,
where $i_{n}$ is the current flowing in the n{\em th} resistor, $N_{neig}$ the
number of nearest neighbors over which the summation is performed. The
parameter $A$ represents the thermal resistance of each resistor and  
determines the importance of Joule heating effects. By taking the above 
expression for $T_n$ we are neglecting time dependent effects in heat 
diffusion studied by Sornette et al. \cite{sornette}. A constant stress 
current $I$ is then applied through perfectly conducting bars at the left and 
right sides of the network.
The two biased processes consist of stochastic transitions between the two 
possible states of each resistor and they occur, through thermal activation,
with probabilities: $W_{Dn}=exp[- E_D/k_B T_n]$ and 
$W_{Rn}=exp[- E_R/k_B T_n]$, characterized by the two activation energies, 
$E_D$ and $E_R$ \cite{SNUBP,prl_stat} ($k_B$ is the Boltzmann constant). 
The network time evolution is obtained by a Monte Carlo simulation which 
updates the network resistance after breaking and recovery processes, 
according to the iterative procedure described in details in Ref. 
\cite{SNUBP}. The sequence of successive configurations provides a resistance 
$R(t)$ signal after an appropriate calibration of the time scale. Then, 
depending on the stress conditions ($I$ and $T_0$) and on the network 
parameters (sizes, activation energies and other parameters related to the 
material like $r_0$ and $\alpha$), the network reaches a steady state or 
undergoes an irreversible electrical failure. This last possibility 
is associated with the achievement of the percolation threshold, $p_c$, for 
the fraction of broken resistors. Therefore, for a given network at a given 
temperature, a threshold current value, $I_B$, exists above which electrical 
breakdown occurs \cite{SNUBP}. Below this threshold, the steady state of the 
network is characterized by fluctuations of the fraction of broken resistors, 
$\delta p$, and of the resistance, $\delta R$, around their respective average
values $<p>$ and $<R>$. In particular, we underline that the ratio 
$(E_D -E_R)/k_B T_0$ determines the average fraction of defects for a given 
value of $I<I_B$ and thus the level of disorder inside the network. 
In the following section we analyze the results of simulations carried out 
by considering networks of different sizes, with different levels of disorder 
and stressed by different currents. In all cases we take $T_0=300$ (K), 
$E_D = 0.170$ (eV), $r_0=1$ ($\Omega$), $\alpha = 10^{-3}$ (K$^{-1}$), 
$A=5 \times 10^5$ (K/W) (these values are chosen as physically plausible). $N$
ranges between $50 \div 125$, while $E_R$ between $0.0259 \div 0.164$ (eV). 

\vspace{-0.6 cm}
\section{Results}
\label{4}
\vspace{-0.6 cm}
We report in Fig. 1 the resistance of a $75 \times 75$ network as a function
of time and at increasing currents. The evolutions $R(t)$ in Fig. 1 
are obtained by taking $E_R = 0.103$ (eV) for the activation energy of the
recovery process. This value of $E_R$ provides a network with an intermediate 
level of disorder. This figure displays two important features of the 
electrical response of a conducting film: i) the linear regime at low currents 
is followed by a nonlinear regime where the average resistance increases
significantly with the current; ii) the amplitude of the resistance 
fluctuations increases strongly close to the breakdown. Precisely, the lowest 
curve in Fig. 1 corresponds to the linear regime, i.e. it is obtained for 
$I=0.001 (A) < I_0$ where $I_0$ is the current value associated with the onset
of the nonlinearity of the I-V characteristic \cite{SNUBP}. The second curve 
corresponds to the nonlinear regime, $I=0.45 (A) > I_0$, the third to the 
threshold for the breakdown, $I=I_B= 0.70 (A)$, and finally the highest curve 
to a network undergoing an electrical breakdown at a current $I=0.75 (A)>I_B$.
This last curve has been shifted upwards by $0.1 (\Omega)$ for graphical 
reasons. A detailed analysis of the behavior of the average resistance and of 
the relative variance of resistance fluctuations as a function of the
current can be found in Ref. \cite{SNUBP}.

  \begin{figure}[bth]
  \centerline{    
  \epsfig{figure=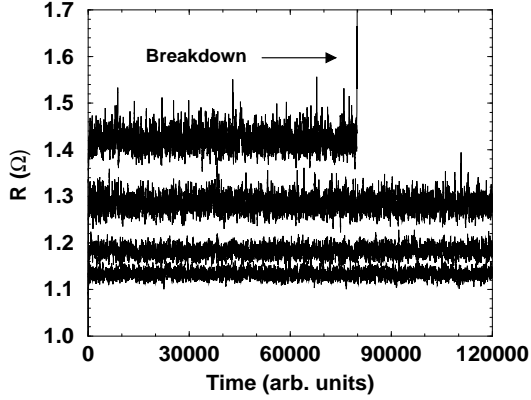,width=0.5\linewidth,height=0.39\linewidth}}
  \vspace{-0.2cm}
  \caption{\small Resistance evolutions at increasing currents. 
Starting from the bottom: $I=0.001$ (A) (linear regime), $I=0.45$ A) 
(nonlinear regime), $I=0.70$ (A) (threshold current), $I=0.75$ (A) 
(breakdown). The highest curve has been shifted upwards by $0.1 (\Omega)$ 
for graphical reasons.}
  \label{fig:1}
  \end{figure}

  \begin{figure}[bth]
  \centerline{    
  \epsfig{figure=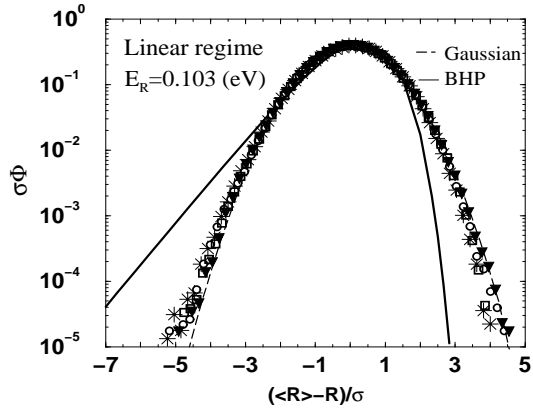,width=0.5\linewidth,height=0.39\linewidth}}
  \vspace{-0.2cm}
  \caption{\small Normalized PDF of resistance fluctuations in the 
linear regime at increasing network sizes: $50\times50$ (stars), 
$75\times75$ (small circles), $100\times100$ (squares), $125\times125$ 
(down triangles). The recovery energy is $E_R=0.103$ (eV). The thick solid 
curve and the dashed curve correspond to the BHP and Gaussian distributions, 
respectively.}
  \label{fig:2}
  \end{figure}

  \begin{figure}[bth]
  \centerline{    
  \epsfig{figure=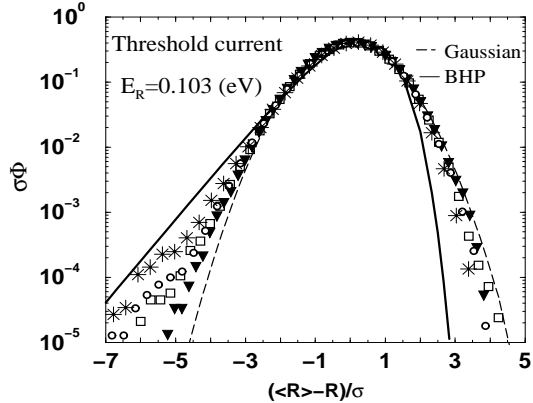,width=0.5\linewidth,height=0.39\linewidth}}
  \vspace{-0.2cm}
  \caption{\small Normalized PDF of resistance fluctuations at the threshold 
current, $I_B$, at increasing network sizes: $50\times50$ (stars), 
$75\times75$ (small circles), $100\times100$ (squares), $125\times125$ 
(down triangles). The recovery energy $E_R$ is the same of Fig. 2. The thick 
solid and the dashed curves also have the same meaning.}
  \label{fig:3}
  \end{figure}

\vspace*{-0.2cm}
We have investigated the Gaussianity of the steady state signals shown 
in Fig. 1. The results of the analysis are reported in Fig. 2 for $I=0.001$ 
(A) (linear regime) and in Fig. 3 for $I=I_B=0.70$ (A) (threshold current for 
breakdown). Precisely, in these figures we show on a lin-log plot the product 
$\sigma \Phi$ as a function of $(<R>-R)/\sigma$, where  $\Phi$ is the 
probability density function (PDF) of the distribution of $\delta R$ and 
$\sigma$ is the root mean square deviation from the average. This normalized 
representation, by making the distribution independent of its first and second
moments, is particularly convenient to  explore the functional form of the 
distribution. These results, together with all the others presented in this
paper have been obtained by considering time series containing about 
$1.2 \times 10^6$ resistance values. Moreover, for comparison, we have also 
reported the Gaussian distribution (which in this normalized representation 
has zero mean and unit variance) and the Bramwell-Holdsworth-Pinton (BHP) 
distribution \cite{bramwell_nat,bramwell_prl}. These authors have proposed a 
universal non-Gaussian PDF for the fluctuations of global quantities of 
systems at criticality\cite{bramwell_nat}. Denoting $m$ a fluctuating quantity
(for example the magnetization of a ferromagnet like in Ref. 
\cite{bramwell_prl}), $<m>$ and $\sigma_m$ its mean value and 
root mean square deviation respectively, $P(m)$ its PDF, 
$y\equiv (m-<m>)/\sigma_m$ the normalized variable, 
$\Pi(y)\equiv \sigma_m P(y)$ the normalized PDF and $x \equiv b(y-s)$, the BHP
distribution takes the following expression \cite{bramwell_prl}: 

\vspace*{-0.15cm}
\begin{equation}
\hspace*{5.cm}\Pi(y) = K [e^{x - e^{x}}]^a \label{eq:bhp}
\end{equation} 
\vspace*{-0.65cm}

where $a=\pi/2$, $b=0.936 \pm 0.002$, $s=0.374 \pm 0.001$ and $K=2.15 \pm0.01$.
This expression can be considered as a generalization of the Gumbel 
distribution, which is often associated with the occurrence of rare events. 
In Figs. 2 and 3 together with the PDF obtained for networks with linear sizes
$N=75$ (small circles) are also displayed the PDF corresponding to $N=50$ 
(stars), $N=100$ (squares) and $N=125$ (down triangles). 
Figure 2 shows that in the linear regime the distribution of $\delta R$ is 
Gaussian for all system sizes. Near breakdown, at $I=I_B$, (Fig. 3) the PDF 
shows non-Gaussian tails. However, this non-Gaussianity progressively 
vanishes when the system size is increased. Therefore, when networks with
intermediate level of disorder are considered, as in the case of Figs. 2 and 
3, deviations from Gaussianity are related to the finite size of the system. 
The PDF in this case is well fitted by the Eq.~(\ref{eq:bhp}) 
once the parameters $a$, $b$, $s$ and $K$ are considered as fitting 
parameters, as described in Ref. \cite{pen_hcis}. Such phenomenological fits 
can provide helpful tools in the study of failure precursors in the case of 
finite size systems.    

  \begin{figure}[bth]
  \centerline{    
  \epsfig{figure=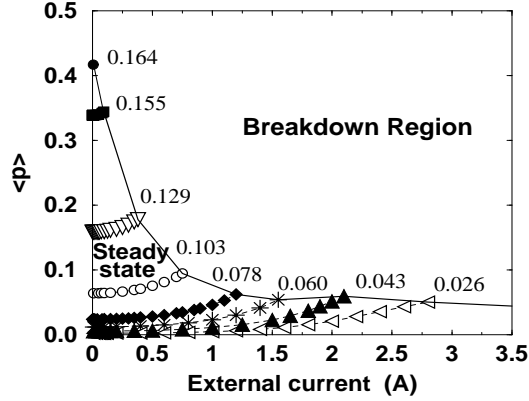,width=0.5\linewidth,height=0.39\linewidth}}
  \vspace{-0.2cm}
  \caption{\small Average fraction of defects versus current in steady-state
networks for different values of the recovery energy $E_R$. The numbers at the
top of each curves indicate $E_R$ in eV. In the region above the 
solid curve the system breaks down.}
  \label{fig:4}
  \end{figure}
  \vspace{-0.15cm}

  \begin{figure}[bth]
  \centerline{    
  \epsfig{figure=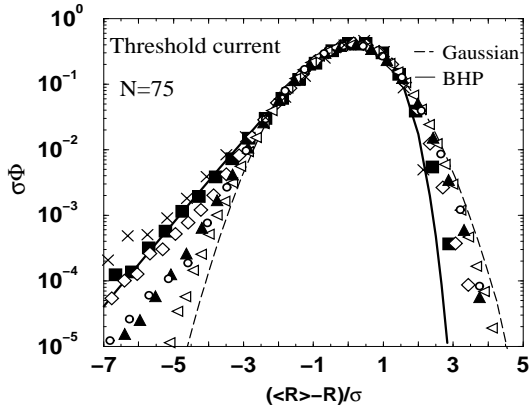,width=0.5\linewidth,height=0.39\linewidth}}
  \vspace{-0.4cm}
  \caption{\small Normalized PDF of resistance fluctuations of a
$75\times75$ network at increasing  $E_R$ values. In all 
cases the current corresponds to the threshold value. Precisely,
left triangles: $E_R=0.026$, $I_B=2.6$; full triangles: $E_R=0.043$, $I_B=2.1$;
small circles: $E_R=0.103$, $I_B=0.70$; diamonds: $E_R=0.140$, $I_B=0.25$; 
full squares: $E_R=0.155$, $I_B=0.085$, crosses:  $E_R=0.164$, $I_B=0.0065$ 
(energies in eV, currents in A).}
  \label{fig:5}
  \end{figure}

  \begin{figure}[bth]
  \centerline{    
  \epsfig{figure=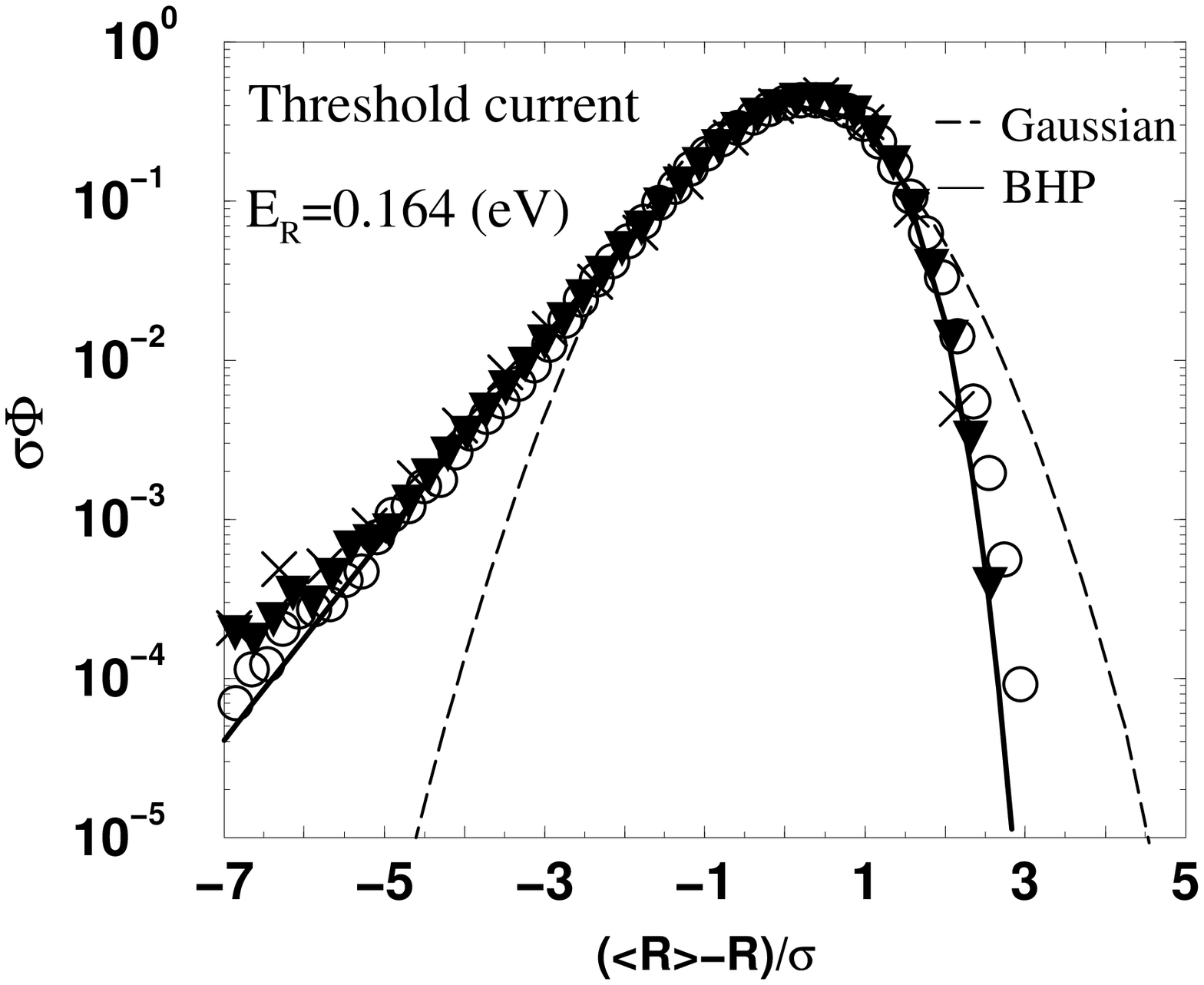,width=0.5\linewidth,height=0.39\linewidth}}
  \vspace{-0.2cm}
  \caption{\small Normalized PDF of resistance fluctuations at the threshold 
current, $I_B$, at increasing network sizes: $75\times75$ (crosses), 
$100\times 100$ (down triangles), $125\times125$ (full circles). The recovery 
energy is $E_R=0.164$ (eV).}
  \label{fig:6}
  \end{figure}
\vspace*{-0.3cm}
We have also considered the role of disorder in the breakdown process and 
its effect on the distribution of resistance fluctuations. At given values of 
$E_D$ and $T_0$, the parameter controlling the disorder inside the network, 
i.e. the average fraction of broken resistors $<p>$, is the activation energy
$E_R$. Figure 4 shows the average fraction of broken resistors versus the 
external current for different values of $E_R$. The data are obtained for a 
$75 \times 75$ network. The different sets of symbols refer to steady 
states of the network. The region above the solid curve corresponds to 
electrical breakdown (in this region the stable states of 
the network are non-conducting states). Thus, Fig. 4 represents the phase 
diagram of the system in the $<p> - I$ plane. It can be shown \cite{upon02}
that all the curves in Fig. 4 collapse onto a single one after normalization 
of $I$ to $I_0$ and of $<p>$ to $<p>_0$ (the defect fraction in the limit of
vanishing current). Moreover, the relative variation of the defect fraction, 
$[<p>-<p>_0]/<p>_0$, scales as the ratio $I/I_0$ with a quadratic exponent 
\cite{SNUBP}. Furthermore, we note that the electrical breakdown
corresponding to the values of $E_R$ considered in Fig. 4, is associated
with a first order phase transition \cite{upon02}. These results agree 
with the behavior observed in electrical breakdown experiments, performed in 
the Joule regime of composites \cite{bardhan}. Nevertheless, it can be shown 
\cite{upon02} that when $E_R$ reaches its maximum value, $E_{R,MAX}$, 
(determined by the stability condition obtained for a system of given size 
in the vanishing current limit \cite{prl_stat,upon02}), the 
conductor-insulator transition becomes of the second order. This change in 
the nature of the transition, when going from small to high disordered 
systems, also agrees with the predictions of Andersen et al. \cite{sornette}. 
Therefore, we have reported in Fig. 5 the normalized PDF calculated at the
breakdown threshold for different values of $E_R$. Again, a $75 \times 75$ 
network is considered. We can see that the non-Gaussianity of the 
distribution at $I=I_B$ increases significantly when increasing the $E_R$ 
value. However, as already shown in Fig. 3, this non-Gaussianity vanishes in 
the large size limit. Nevertheless, when the value of $E_R$ is nearly equal to 
$E_{R,MAX}$, the system approaches the critical point and the PDF tends to 
achieve the BHP form, independently of system size. This is shown in Fig. 6 
which displays the normalized PDF calculated at the breakdown threshold, 
$I=I_B$, for $E_R=0.164$ (eV). The three sets of data are obtained for $N=75$,
$I=0.0065$ (A) (crosses), $N=100$ and $I=0.009$ (A) (down triangles) and 
$N=125$ and $I=0.011$ (A) (full circles) and correspond to networks near their
critical point (the data for $N=50$ are not shown in Fig. 6 because 
$50 \times 50$ networks are always unstable for this value of $E_R$). 
We can see that in this case the PDF becomes independent of the sizes
of the system and it is well fitted by the BHP distribution. 

\vspace{-1.1 cm}
\section{Conclusions}
\label{5}
\vspace{-1.0 cm}
We have studied the distribution of the resistance fluctuations of conducting 
thin films with granular structure near electrical breakdown.
The study has been performed by describing the film as a resistor network
and by using the SNUBP model \cite{SNUBP}. We have considered systems of 
different sizes and with different levels of internal disorder. A 
non-Gaussianity of the fluctuation distribution is found near electrical 
breakdown. This non-Gaussianity increases with the degree of disorder  
of the network. However, we have shown that this non-Gaussianity
is related to the finite size of the system and that it vanishes in the 
large size limit. Nevertheless, near the critical point the non-Gaussianity 
persists in the large size limit and is well fitted by the universal 
Bramwell-Holdsworth-Pinton distribution 
\cite{bramwell_nat,bramwell_prl,bramwell_pre}.

\end{document}